\documentclass[sigconf]{acmart}

\AtBeginDocument{%
  }
\usepackage[utf8]{inputenc}
\usepackage[T1]{fontenc}
\usepackage{hyperref}
\usepackage{url}
\usepackage{booktabs}
\usepackage{amsfonts}
\usepackage{nicefrac}
\usepackage{microtype}
\usepackage{xcolor}

\usepackage{wrapfig}

\usepackage{multirow, booktabs}
\usepackage{longtable}
\usepackage{makecell}
\usepackage{pifont}
\usepackage{graphicx}
\usepackage{tabularx}
\usepackage{amsmath}
\usepackage{amsfonts}  
\usepackage{mathrsfs}
\usepackage{amsthm}
\usepackage{hyperref}
\usepackage{fancyhdr}
\usepackage{mdframed}
\usepackage{caption}
\usepackage[misc]{ifsym}
\definecolor{customblue}{RGB}{199,234,250}
\newcommand{\name}{\texttt{BurstGPT}}
\newcommand{\daysoftrace}{213} %{121}
\begin{document}

\title{BurstGPT: A Real-World Workload Dataset to \\ Optimize LLM Serving Systems}
% \author{Yuxin Wang$^{1*}$, Yuhan Chen$^{1*}$, Zeyu Li$^1$, Xueze Kang$^1$, Yuchu Fang$^5$, Yeju Zhou$^5$, Zheng Yang$^5$, Zhenheng Tang$^1$, Xin He$^5$, Rui Guo$^3$, Xin Wang$^3$, Qiang Wang$^2$, Amelie Chi Zhou$^4$, \\\Letter \ Xiaowen Chu$^1$}
% \affiliation{%
%   \institution{The Hong Kong University of Science and Technology (Guangzhou)$^{1}$, Tsinghua University$^{3}$, \\Harbin Institute of Technology (Shenzhen)$^{2}$, \\Hong Kong Baptist University$^{4}$, Huawei Technologies Co.$^{5}$}
%   }
% \email{xwchu@hkust-gz.edu.cn}

\author{Yuxin Wang}
\authornote{Both authors contributed equally to this research.}
\affiliation{%
  \institution{HKRC}
  \city{Hong Kong}
  \country{CHINA}
}

\author{Yuhan Chen}
\authornotemark[1]
\affiliation{%
  \institution{HKUST (Guangzhou)}
  \city{Guangzhou}
  \state{Guangdong}
  \country{CHINA}
}

\author{Zeyu Li}
\affiliation{%
  \institution{HKUST (Guangzhou)}
  \city{Guangzhou}
  \state{Guangdong}
  \country{CHINA}
}

\author{Xueze Kang}
\affiliation{%
  \institution{HKUST (Guangzhou)}
  \city{Guangzhou}
  \state{Guangdong}
  \country{CHINA}
}

\author{Yuchu Fang}
\affiliation{%
  \institution{Huawei Technologies Co.}
  \city{Shenzhen}
  \state{Guangdong}
  \country{CHINA}
}

\author{Yeju Zhou}
\affiliation{%
  \institution{Huawei Technologies Co.}
  \city{Shenzhen}
  \state{Guangdong}
  \country{CHINA}
}

\author{Yang Zheng}
\affiliation{%
  \institution{Huawei Technologies Co.}
  \city{Shenzhen}
  \state{Guangdong}
  \country{CHINA}
}

\author{Zhenheng Tang}
\affiliation{%
  \institution{HKUST}
  \state{Hong Kong}
  \country{CHINA}
}

\author{Xin He}
\affiliation{%
  \institution{Hong Kong Baptist University}
  \state{Hong Kong}
  \country{CHINA}
}

\author{Rui Guo}
\affiliation{%
  \institution{Tsinghua University}
  % \city{Beijing}
  \state{Beijing}
  \country{CHINA}
}

\author{Xin Wang}
\affiliation{%
  \institution{Tsinghua University}
  \state{Beijing}
  \country{CHINA}
}

\author{Qiang Wang}
\affiliation{%
  \institution{HIT (Shenzhen)}
  \city{Shenzhen}
  \state{Guangdong}
  \country{CHINA}
}

\author{Amelie Chi Zhou}
\affiliation{%
  \institution{Hong Kong Baptist University}
  % \city{Guangzhou}
  \state{Hong Kong}
  \country{CHINA}
}

\author{Xiaowen Chu}
% \authornotemark[2]
\authornote{Corresponding author. Email: xwchu@hkust-gz.edu.cn.}
% \email{xwchu@hkust-gz.edu.cn}
\affiliation{%
  \institution{HKUST (Guangzhou)}
  \city{Guangzhou}
  \state{Guangdong}
  \country{CHINA}
}


\begin{abstract}
Serving systems for Large Language Models (LLMs) are often optimized to improve quality of service (QoS) and throughput. However, due to the lack of open-source LLM serving workloads, these systems are frequently evaluated under unrealistic workload assumptions. Consequently, performance may degrade when systems are deployed in real-world scenarios.

This work presents {\name}, an LLM serving workload with 10.31 million traces from regional Azure OpenAI GPT services over {\daysoftrace} days. 
{\name} captures LLM serving characteristics from user, model and system perspectives:
(1) User request concurrency: burstiness variations of requests in Azure OpenAI GPT services, revealing diversified concurrency patterns in different services and model types.
(2) User conversation patterns: counts and intervals within conversations for service optimizations.
(3) Model response lengths: auto-regressive serving processes of GPT models, showing statistical relations between requests and their responses.
(4) System response failures: failures of conversation and API services, showing intensive resource needs and limited availability of LLM services in Azure.
The details of the characteristics can serve multiple purposes in LLM serving optimizations, such as system evaluation and trace provisioning. In our demo evaluation with \name, frequent variations in \name~reveal declines in efficiency, stability, or reliability in realistic LLM serving. We identify that the generalization of KV cache management, scheduling and disaggregation optimizations can be improved under realistic workload evaluations.
\name~is publicly available now at \url{https://github.com/HPMLL/BurstGPT} and is widely used to develop prototypes of LLM serving frameworks in the industry.
\end{abstract}


\maketitle

\section{Introduction}
Large language models (LLMs), particularly Generative Pre-trained Transformer (GPT) models, have significantly impacted the AI industry. Notable examples include OpenAI's ChatGPT~\cite{chatgpt}\cite{openai2023gpt4}, Google's Gemini~\cite{geminiteam2024gemini}, and Meta's Llama~\cite{touvron2023llama}. However, deploying and operating LLM services is costly due to the substantial use of AI accelerators~\cite{li2023alpaserve}. The high volume of user requests and limited hardware resources further constrain the quality of service (QoS)~\cite{li2023alpaserve}. Therefore, optimizing serving systems to reduce costs and improve user experience is urgent.

\begin{figure}[t]
    \centering
    \includegraphics[width=0.48\textwidth]{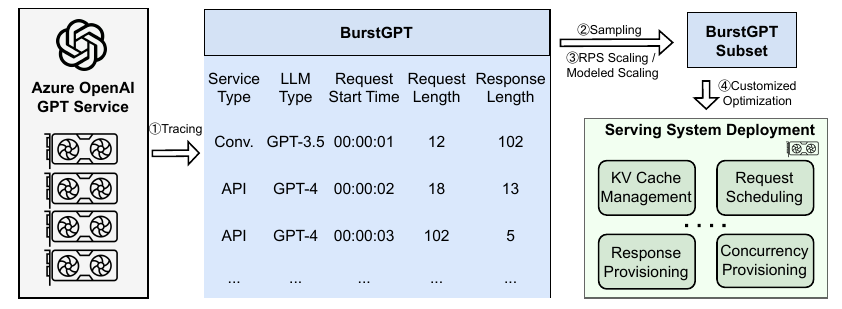}
    \caption{Data collection and use method of \name. \name~is a real-world workload trace from the Azure OpenAI GPT service. A scaled sample from a period of \name~can be used to optimize serving systems using specific methods, considering realistic concurrency and response patterns. Note that we open-sourced two versions of \name: a cleaned trace and a raw trace, with failure logs excluded from the cleaned version.}
    \label{fig:pipeline}
\end{figure}

Existing systems primarily optimize system throughput and achieve better QoS. However, these studies often use non-LLM serving workloads in evaluations, derived from synthetic models~\cite{wu2023fast} or non-LLM services such as Microsoft Azure Functions (MAF)~\cite{maf}\cite{maf2}. This work identifies that these unrealistic synthetic workloads do not reflect the workload patterns of auto-regressive LLM serving and limit the effectiveness of evaluating these systems.

In addressing this gap, we introduce a real-world LLM serving workload dataset named {\name}. As shown in Figure \ref{fig:pipeline}, the traces are collected from the Azure OpenAI GPT Service~\cite{azureopenai}, which serves GPT models on Azure and provides APIs for regional GPT service providers, such as enterprises and campuses, to build customized GPT services.
We collected data from one of these regional GPT service providers with over 3,000 users. By deploying a logging engine on this regional service provider, we monitored privacy-independent metrics of each GPT request. In Figure \ref{fig:pipeline}, we collected the \textit{request start time}, \textit{request and response token lengths}, \textit{service type} (API or conversational), and \textit{LLM type} (ChatGPT, GPT-4 or GPT-4v) for {\daysoftrace} days. 
The dataset comprises 8.69 million traces of ChatGPT and 0.95 million GPT-4 traces using API services, and 0.30 million ChatGPT traces and 0.16 million traces of GPT-4 using conversational services. It also includes failure requests while the response length is zero.

To use \name, the user can sample a period of trace and scale it to fit the target system size. Users can apply vanilla \name~using request-per-second (RPS) scaling or model the trace and adjust parameters, as shown in Figure \ref{fig:pipeline}. The subset can be used for various optimizations, such as identifying concurrency patterns for provisioning and predicting response lengths based on request-response length distributions. 
Additionally, this paper provides a detailed demo of \name~for LLM serving evaluation. Key findings from our analysis and evaluations are listed below:
\begin{itemize}
    \item{User Request Concurrency.} We identify unique patterns in request concurrency and lengths, varying by service types. This insight suggests that service-specific user request pattern evaluations are important for optimizing LLM serving. In our evaluation, (1) concurrency from non-LLM workloads does not accurately reflect system performance as \name~does. (2) Even for LLM workloads, generalization of system optimization across serving types is not guaranteed. For example, we verify that while optimized first-come-first-served scheduling is effective for conversation services, its efficiency can be reduced for API services.

    \item{User Conversation Patterns.}
    We share analyze traces for the first 2 months in \ref{subsec_con}, \ref{subsec_len}, \ref{subsec_fail} and conversation traces from the last 3 months in \ref{subsec_conv}. The detailed analysis of conversation traces can help schedule LLM serving, e.g., KV cache offload strategies.
    
    \item{Model Response Lengths.} Longer response lengths lead to higher workloads during LLM serving. The unpredictable nature of response lengths per request introduces uncertainty to system pressure. The statistical study of the relationship between request and response lengths in \name~\\ sheds light on future response provisioning, which can help improve system performance.

    \item{System Response Failures.} We observe a relatively high failure rate in \name. In our evaluation, we identify that this is primarily due to inefficiencies in KV cache management. Variations in burstiness in \name~lead to memory bottlenecks, causing spikes in failure rates and performance degradation.
\end{itemize}

With the dataset, we open-sourced a benchmark suite, BurstGPT-Perf. In our evaluation using BurstGPT-Perf, we first present our insights on using \name~to assess system performance differences between LLM and non-LLM workloads. We also demonstrate use cases of \name~in scheduling policy selection, workload provisioning. Then, we share our insights in prefill-decode(PD) disaggregation in industrial system prototypes. The evaluation results show that \name~can effectively suggest optimizations for LLM serving systems in real-world scenarios.

\section{Preliminary and Motivation}\label{sec:back}

\begin{table}[!hb]
  \centering
  \caption{ Comparison of BurstGPT with Other Traces}
  % \vspace{-6pt}
      \setlength{\aboverulesep}{0pt}
  \setlength{\belowrulesep}{0pt}
  \setlength{\extrarowheight}{0pt}
  \scalebox{0.8}{
      \begin{tabular}{
    |p{1.1in}| % Workloads column
    >{\centering\arraybackslash}m{0.5in}| % {\name} column
    >{\centering\arraybackslash}m{0.5in}| % MAF1 column
    >{\centering\arraybackslash}m{0.5in}| % MAF2 column
    >{\centering\arraybackslash}m{0.65in}| % FastServe column
  }
    \toprule
    \multicolumn{1}{|c|}{Serving} & \multicolumn{3}{c|}{Real-world} & Synthetic \\
\cmidrule{2-5}    \multicolumn{1}{|c|}{Workloads} & {\name} & MAF1~\cite{maf}  & MAF2~\cite{maf2}  & FastServe~\cite{wu2023fast} \\
    \midrule
    \midrule

    % Real-world Requests & \ding{51}     & \ding{53}     & \ding{53}     & \ding{53} \\
% \cmidrule{1-1}    

    LLM-User Req. Pat. & \ding{51}     & \ding{53}     & \ding{53}     & \ding{53} \\
    \midrule
    LLM-User Conv. Pat. & \ding{51}     & \ding{53}     & \ding{53}     & \ding{53} \\
        \midrule
    LLM Resp. Len. & \ding{51}     & \ding{53}     & \ding{53}     & \ding{51} \\
    \midrule
    Sys. Resp. Fail.  & \ding{51}     & \ding{53}     & \ding{53}     & \ding{53} \\
    \bottomrule
    \end{tabular}}
  \label{tab:sum}
\end{table}%

\subsection{Limitations of LLM Serving}
Each request in LLM serving requires uncertain but large computational due to the model size and auto-regressive nature~\cite{kwon2023efficient}. To reduce deployment and operational expenses, several specialized frameworks have been developed, including TensorRT-LLM~\cite{trt}, vLLM~\cite{kwon2023efficient}, DeepSpeed~\cite{rasley2020deepspeed}, and lightLLM~\cite{lightllm}. In addition to these frameworks, various memory and compute optimizations have been proposed for efficient decoding~\cite{flashdecoding}\cite{hong2023flashdecoding++}\cite{ge2023model}\cite{xiao2023smoothquant}\cite{dettmers2022llm}\cite{shazeer2019fast}\cite{ainslie2023gqa}\cite{zhou2024training}\cite{tang2024fusefl}, significantly improving LLM inference performance in deployments.
Also, efficient request scheduling is crucial for optimizing LLM service. For example, Orca~\cite{orca} introduces iteration-level scheduling, that dynamically adjusts batch size during iterations; FastServe~\cite{wu2023fast} proposes optimized Multilevel-Feedback-Queue (MLFQ) scheduling in the request queue to improve system efficiency.

However, these optimizations have yet to be evaluated on real-world LLM serving workloads. We identify that generalizing serving optimizations from non-LLM or synthetic workloads to LLM workloads can be challenging~\cite{miao2023efficient}\cite{miao2023specinfer} for ensuring serving efficiency, stability, and reliability.

\subsection{Towards Workload-aware LLM Serving}
Stable workloads lead to consistent system performance across various metrics. However, in the real world, the workload of LLM is diversified due to user, system, and model behaviors. 
Neglecting any aspect of workload traces during system optimization may yield an incomplete understanding of a framework's behavior in real-world deployment. 

\paragraph{User Concurrency Patterns}
Currently, the assessment of performance in LLM serving frameworks relies on synthetic concurrency~\cite{wu2023fast} or non-LLM concurrency, such as Microsoft Azure Functions (MAF)~\cite{maf}\cite{maf2}. As concluded in Table \ref{tab:sum}, these approaches do not reflect real-world LLM workloads due to a lack of empirical data on system and model patterns, as well as user behaviors. For example, the request-per-second (RPS) metric in MAF is much higher (1.64 RPS on average) compared to LLM services (0.019 RPS on average in conversation service and 0.21 RPS on average in API service when using ChatGPT) due to the non-autoregressive and lightweight nature of function services in Azure. However, burstiness in concurrency is much less severe in MAF, likely leading to performance divergence from LLM workloads.

\paragraph{Model Patterns}
The computational complexities of prefilling and decoding in LLM systems are $O(s^2)$ and $O(s)$ relative to the request length $s$, respectively~\cite{vllm}. This relationship between request length and system load significantly differs from that observed in other lightweight cloud functions and synthetic workloads~\cite{maf}\cite{maf2}, highlighting a unique resource occupation of LLM serving. 
These complexities are also reflected in the request concurrency, failure rates, and latency of each request in LLM serving.

\section{Introduction to BurstGPT}\label{sec:trace}
\begin{figure}[!h]
    \centering
    \begin{minipage}{0.35\textwidth}
        \centering
        \includegraphics[width=\textwidth]{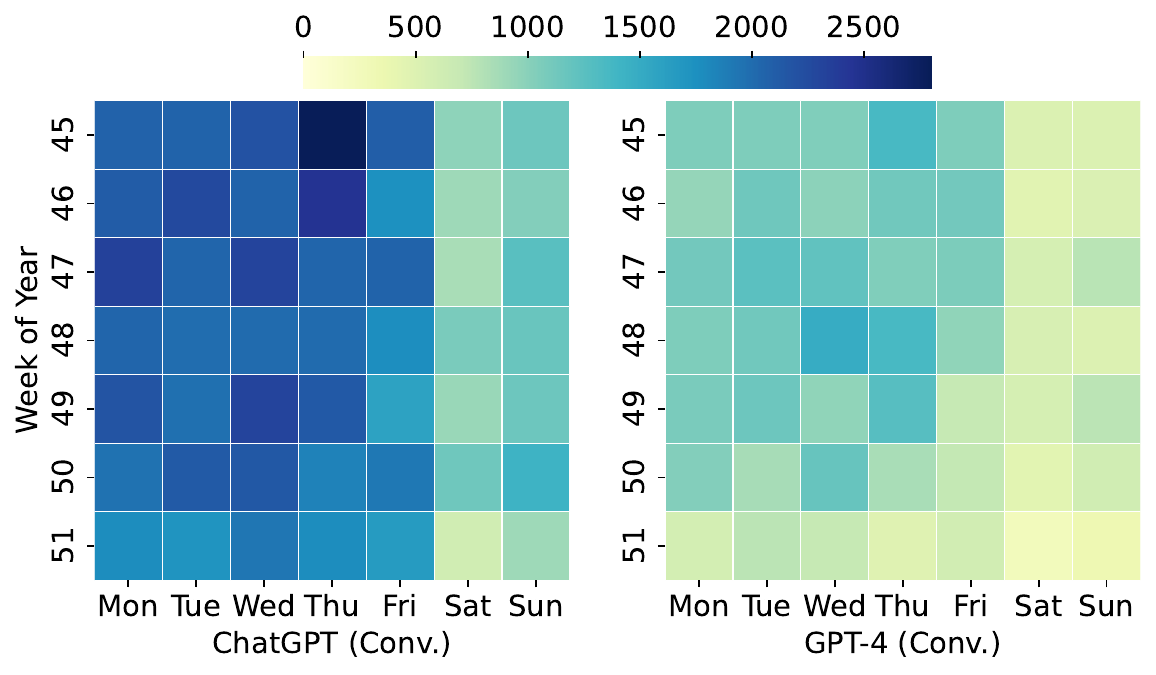}
        \vspace{-18pt}
        \centering
        \captionsetup{font=footnotesize}
        \caption{Weekly Periodicity of Conversation Services in \name.}
        \label{fig:periodic1}
    \end{minipage}
    \begin{minipage}{0.35\textwidth}
        \centering
        \includegraphics[width=\textwidth]{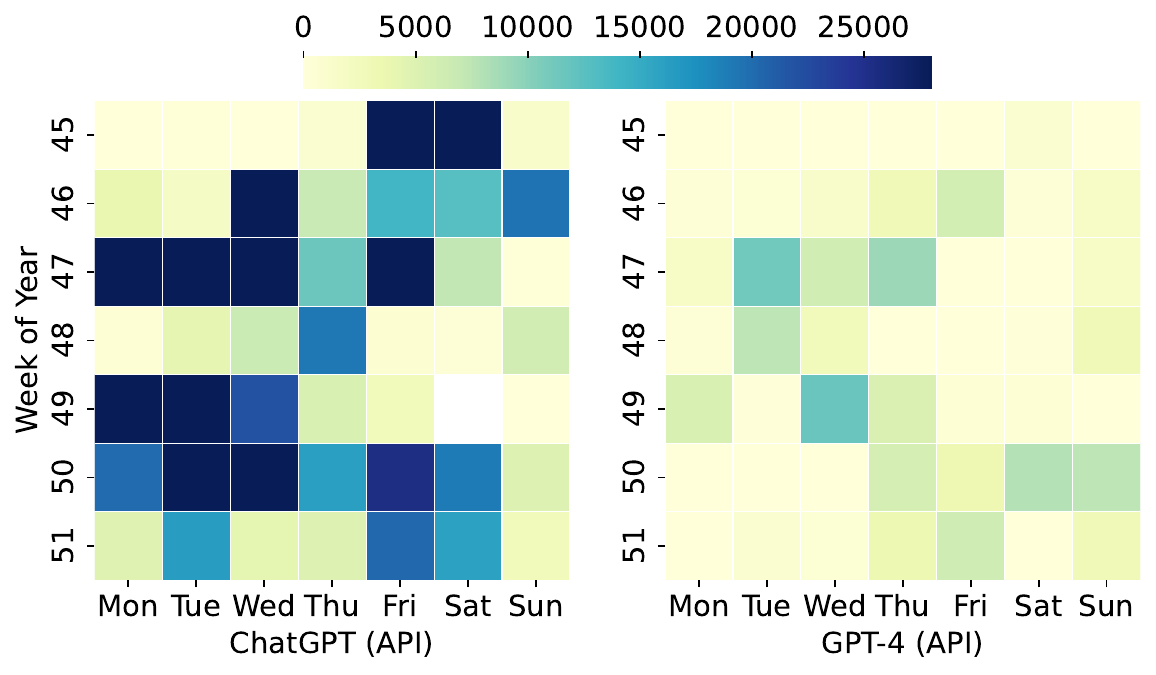}
        \vspace{-18pt}
        \captionsetup{font=footnotesize}
        \caption{Weekly Aperiodicity API Services in \name.}
        \label{fig:burst1}
    \end{minipage} 
\end{figure}

\begin{figure}[!h]
    \centering
    \begin{minipage}{0.39\textwidth}
        \centering
        \includegraphics[width=\textwidth]{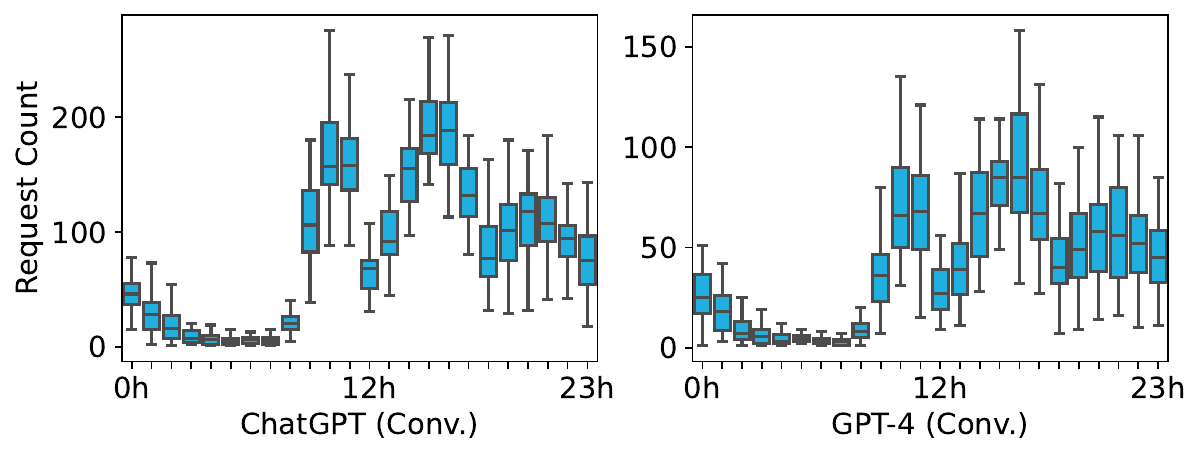}
        \vspace{-18pt}
        \captionsetup{font=footnotesize}
        \caption{Daily Periodicity Conversation Services in \name.}
        \label{fig:periodic2}
    \end{minipage}
    \begin{minipage}{0.39\textwidth}
        \centering
        \includegraphics[width=\textwidth]{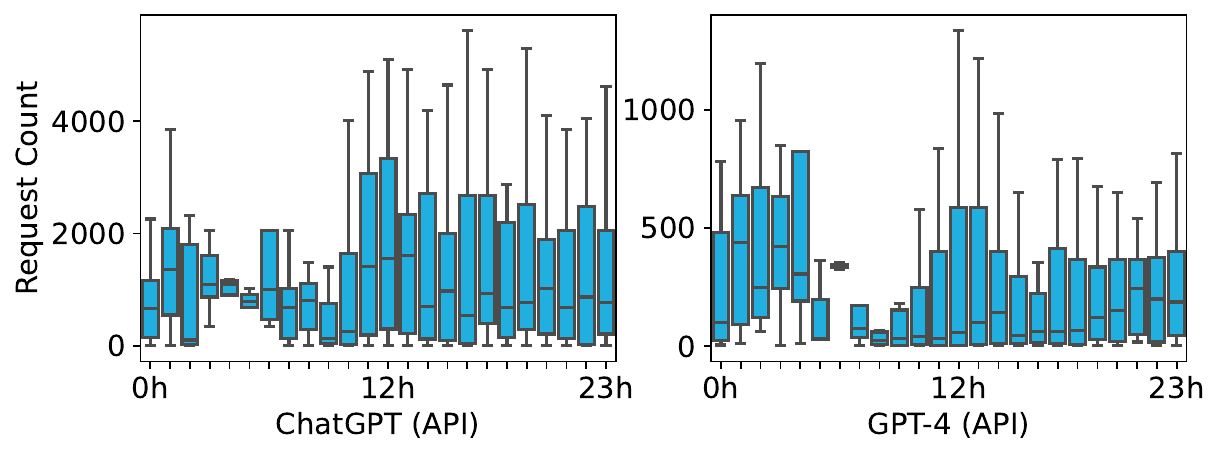}
        \vspace{-18pt}
        \captionsetup{font=footnotesize}
        \caption{Daily Aperiodicity API Services in \name.}
        \label{fig:burst2}
    \end{minipage}
\end{figure}

\subsection{User Request Concurrency}
\label{subsec_con}

\begin{figure}[h]
    \centering
    % \begin{minipage}{0.35\textwidth}

    %     \centering
    %     \includegraphics[width=\textwidth]{figures/burst_factor_cnt.pdf}
    %     \centering
    %     \captionsetup{font=footnotesize}
    %     \caption{Frequency of Doubling in Request Volume During Workdays in \name.}
    %     \label{fig:short1}
    % \end{minipage}
    \begin{minipage}{0.41\textwidth}
        \centering
        \includegraphics[width=\textwidth]{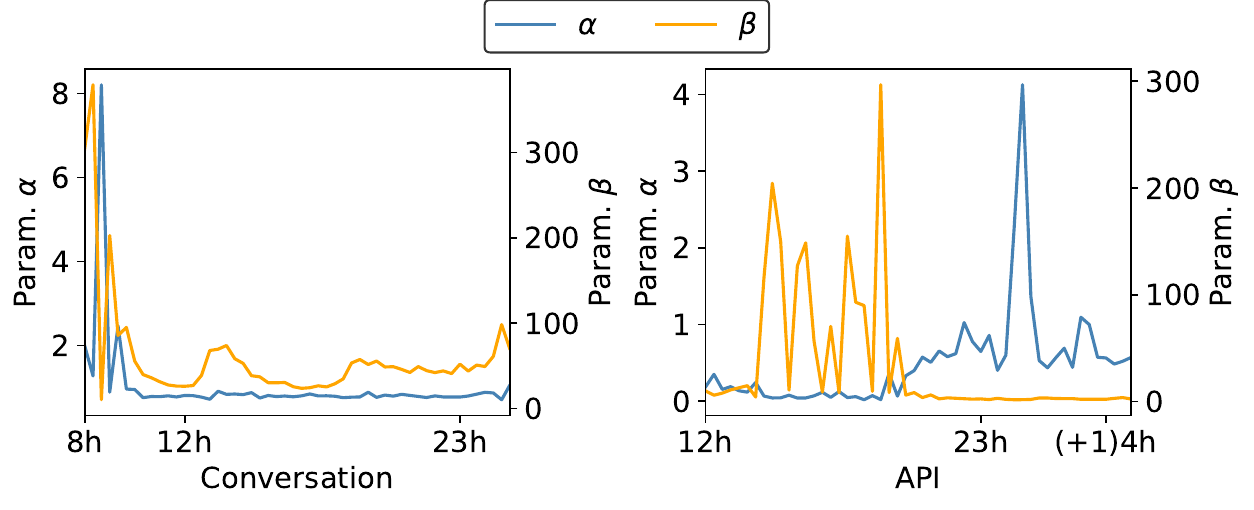}
        \vspace{-10pt}
        \captionsetup{font=footnotesize}
        \caption{Variation of Burstiness in Gamma Distribution in \name.}
        \label{fig:short2}
    \end{minipage}
\end{figure}

\begin{figure}[b]
    \centering
    \begin{minipage}{0.41\textwidth}
        \centering
        \includegraphics[width=\textwidth]{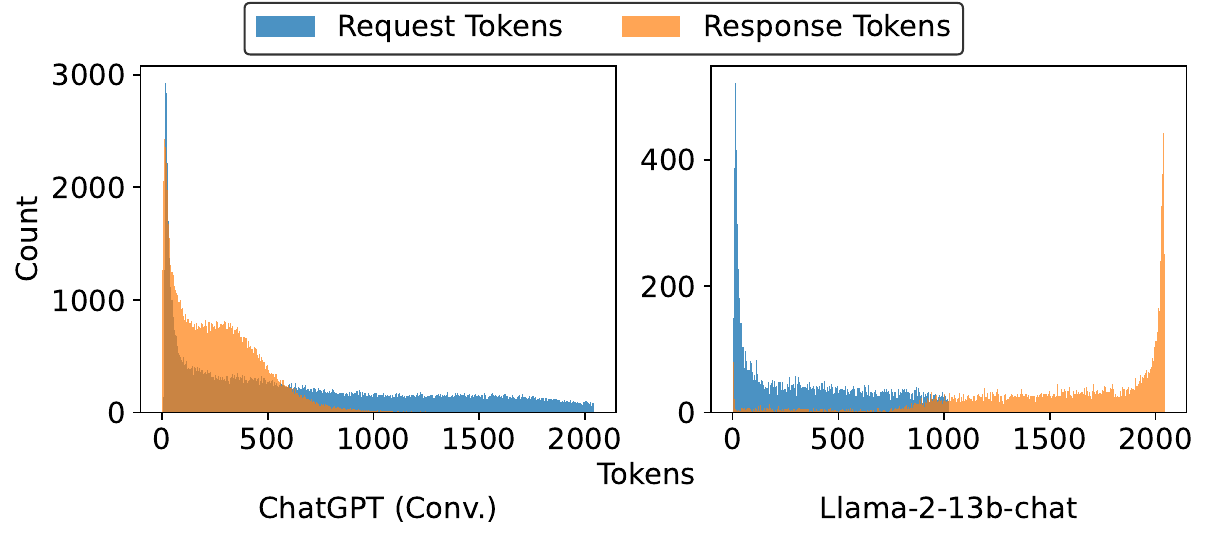}
        % %\vspace{-18pt}
        \centering
        \captionsetup{font=footnotesize}
        \caption{Distribution of Request and Response Tokens.}
        \label{fig:count1}
    \end{minipage}
    \begin{minipage}{0.41\textwidth}
        \centering
        % \vspace{-11pt}
        \includegraphics[width=\textwidth]{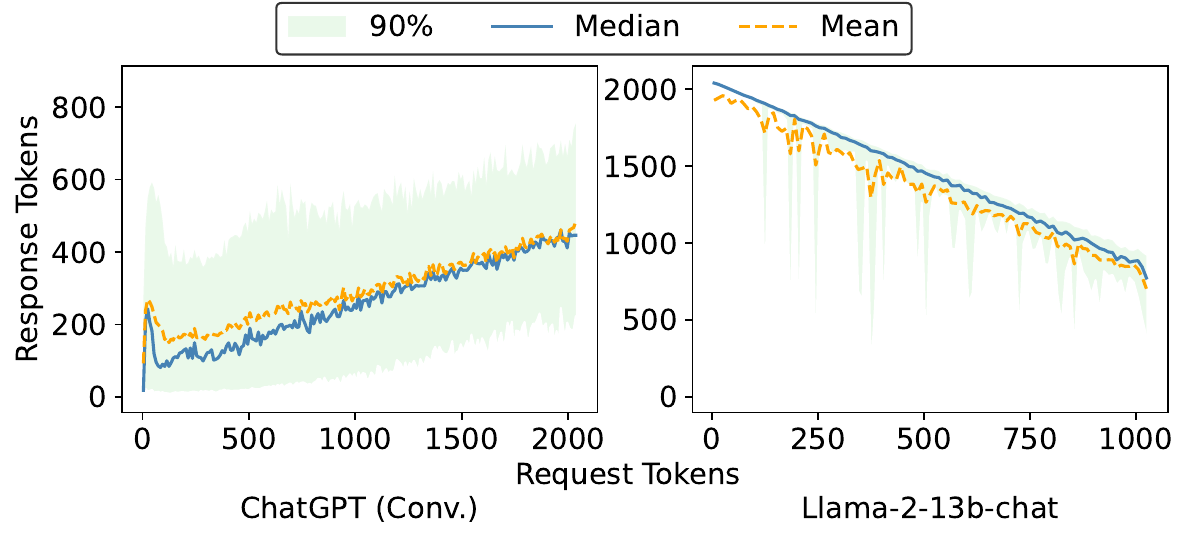}
        \captionsetup{font=footnotesize}
        \caption{Response Lengths per Request}
        \label{fig:count2}
    \end{minipage}
\end{figure}
\paragraph{Long-term Patterns: Periodicity and Aperiodicity}
\label{sec:Long-termPatterns}
% The long-term usage patterns of conversation services, spanning hours and days, exhibit periodic characteristics without distinct burstiness. As illustrated in Figure~\ref{fig:periodic1}, there is a notable trend for both ChatGPT and GPT-4, where conversation volumes peak during weekdays and diminish during weekends. This trend implies heightened user interaction with GPTs, primarily on workdays. Notably, GPT-4's overall usage is less compared to ChatGPT. Also, Figure~\ref{fig:periodic2} demonstrates that the conversation volume for both ChatGPT and GPT-4 exhibits periodic highs during working hours and lows during night hours.

% Contrastingly, the long-term pattern of API services, considering the same hourly and daily intervals, follows an aperiodic pattern characterized by burstiness. Figure~\ref{fig:burst1} and Figure~\ref{fig:burst2} reveal irregular, dense request submissions to API services, diverging significantly from the more predictable patterns observed in conversation services. The variance in daily request volumes, as indicated by a more significant standard deviation compared to Figure~\ref{fig:periodic2}, suggests that these requests may be automated. This irregularity could be attributed to researchers utilizing API services for varied research activities.

The long-term usage patterns of conversation services, spanning hours and days, exhibit periodic characteristics. As illustrated in Figure~\ref{fig:periodic1}, conversation volumes for both ChatGPT and GPT-4 peak during weekdays and diminish during weekends, implying heightened user interaction primarily on workdays. Notably, GPT-4's overall usage is lower compared to ChatGPT. Figure~\ref{fig:periodic2} shows that conversation volumes for both ChatGPT and GPT-4 exhibit periodic highs during working hours and lows during night hours.

In contrast, the long-term pattern of API services, over the same hourly and daily intervals, follows an aperiodic pattern characterized by burstiness. Figures~\ref{fig:burst1} and~\ref{fig:burst2} reveal irregular, dense request submissions to API services, significantly diverging from the more predictable patterns observed in conversation services. The higher variance in daily request volumes, indicated by a greater standard deviation compared to Figure~\ref{fig:periodic2}, suggests these requests may be automated. 
% This irregularity could be attributed to researchers utilizing API services for varied research activities.

\paragraph{Short-term Patterns: Variant Burstiness}

Figure~\ref{fig:short2} presents the modeled burstiness every twenty minutes in ChatGPT's conversation and API services during working hours on weekdays. The burstiness is modeled using the Gamma distribution, as supported in~\cite{li2023alpaserve}\cite{wu2023fast}. 
We observe that the shape parameter $\alpha$ and the rate parameter $\beta$ of the Gamma distribution vary between conversation and API services. A smaller $\alpha$ indicates a higher coefficient of variation (CV) of the Gamma distribution, meaning the workload is more bursty~\cite{wu2023fast}. In conversation services, the parameter $\alpha$ varies sharply during working hours. In API services, both $\alpha$ and $\beta$ vary sharply during the day. These observations underscore the instability of burstiness in the system.

\subsection{User Conversation Patterns } \label{subsec_conv}

\begin{figure}[!h]
\centering
    \begin{minipage}{0.41\textwidth}
        \centering
        \includegraphics[width=\textwidth]{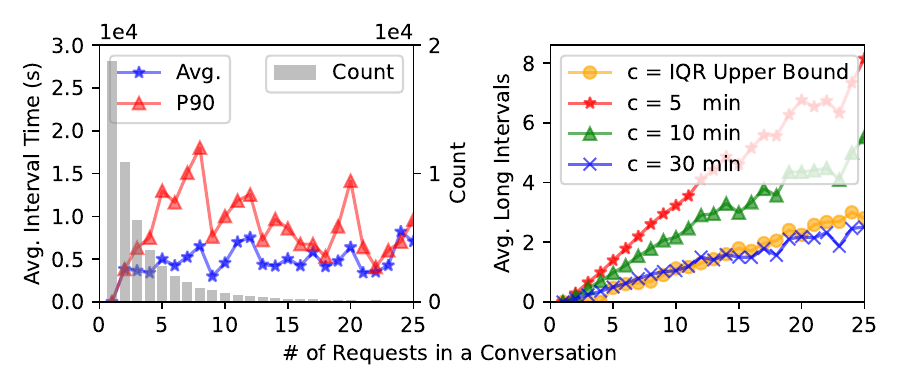}
        \vspace{-10pt}
        \captionsetup{font=footnotesize}
        \caption{Count, Average Interval times, and Average Long Intervals as the Number of Requests in a Conversation in \name.} 
        \label{fig:conv}
    \end{minipage}
\end{figure}

\paragraph{Distribution}
Over 35\% of conversations end with only one request. The left of Figure \ref{fig:conv} shows that the distribution of conversation requests exhibits an exponential decline, with a median of 2 and 75\% of conversations with four or fewer requests.

\paragraph{Interval Time} 
The left of Figure \ref{fig:conv} shows the average interval time is not directly related to the number of requests. The P90 interval time increases when the number of requests is less than 5 and then fluctuates after that.

\paragraph{The More Requests, the Longer Intervals}
The right of Figure \ref{fig:conv} illustrates the number of intervals exceeding a certain duration $c$ in each conversation, which we define as long intervals. We make $c$ equal to 3 different time deltas and the interquartile range (IQR) upper bound. These intervals increase as the number of conversation requests increases.

\subsection{Model Response Patterns} \label{subsec_len}
To identify differences in response length distributions across LLMs, this subsection employs conversational traces in \name~and uses Llama-2-13b-chat\cite{touvron2023llama} for comparison. For Llama-2-13b-chat, we feed 10k real-world prompts to the model for inference and gather the response lengths. The prompts are randomly truncated from the ShareGPT dialogues~\cite{gudibande2023false} with alignined distribution to \name's conversational traces~\cite{zheng2023response}. This alignment allows for comparing the response length distributions between the two LLMs, facilitating a deeper understanding of their behaviors in responding to conversational prompts.

\paragraph{Distribution of Request and Response Lengths}
In Figure~\ref{fig:count1}, we observe similarities in request distributions and differences in response distributions. Both ChatGPT and Llama-2-13b-chat request distributions adhere to a Zipf distribution~\cite{wu2023fast}, characterized by a peak in the frequency of shorter requests.
The histogram of ChatGPT's response tokens reveals a bimodal distribution of length variability. In contrast, Llama-2-13b-chat exhibits a Zipf distribution with a higher frequency of longer requests. These distributions provide insights into the conversational dynamics of request and response lengths for the models. ChatGPT handles longer contexts compared to Llama, reflecting differences in model capacities or interactions.

\paragraph{Response Lengths per Request and Request Bin}
ChatGPT shows a linear correlation between request and response lengths, with a symmetric distribution of response lengths in Figure~\ref{fig:count2}. In contrast, Llama-2-13b-chat decreases response length as request length increases, with higher variability for longer requests due to context length limits. 

\begin{figure}[!h]
    \centering
    \begin{minipage}{0.41\textwidth}
        \centering
        \includegraphics[width=\textwidth]{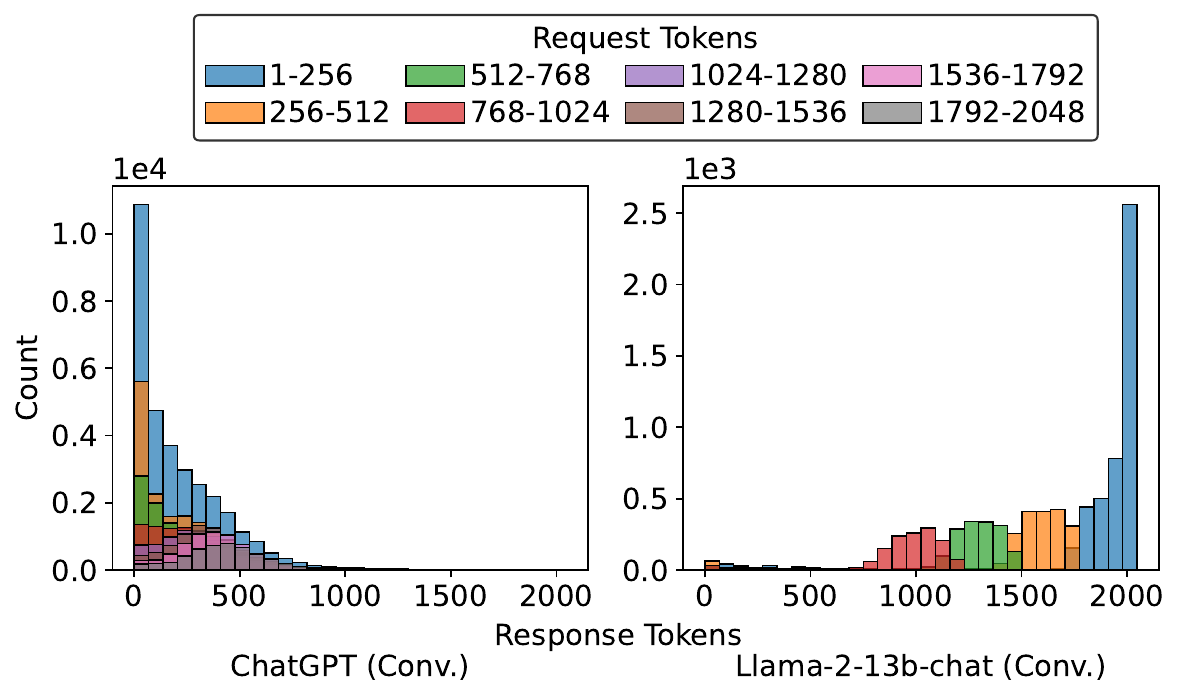}
        \vspace{-10pt}
        \captionsetup{font=footnotesize}
        \caption{Response Lengths per Request Bin}
        \label{fig:count3}
    \end{minipage}
\end{figure}

In Figure~\ref{fig:count3}, ChatGPT tends to produce shorter responses than Llama-2-13b-chat, indicated by a shifted Gaussian distribution with a peak at shorter lengths and a longer spread as request length increases. In contrast, Llama-2-13b-chat shows less variability in response length, peaking at 2048 characters.

% \begin{mdframed}[backgroundcolor=customblue]
% The request lengths' distribution across models is consistent with the \textit{Zipf distribution} in our analysis, contrasted by significant divergence in the response lengths' distributions. The response lengths of models like ChatGPT and Llama-2-13b-chat are different: ChatGPT shows varied response lengths, while Llama tends towards longer, more consistent responses. 

% \end{mdframed}

\subsection{System Response Failures} \label{subsec_fail}
% We observe evident high failure rate of GPT serving in some services.
% Figure~\ref{fig:error} indicates that ChatGPT has a significantly higher service failure rate than GPT-4 in conversational and API interfaces. Overall, GPT-4 improves service reliability over ChatGPT, and demonstrates greater consistency and stability with fewer outliers. Also, the average failure rate of conversation service is consistently high, e.g., over 5\% for ChatGPT service, which is significantly higher than regular cloud services~\cite{ahsan2023failure}. We will discuss the reliability problem in detail in Section~\ref{sec:result}.

\begin{figure}[!h]
\centering
    \begin{minipage}{0.41\textwidth}
        \centering
        \includegraphics[width=\textwidth]{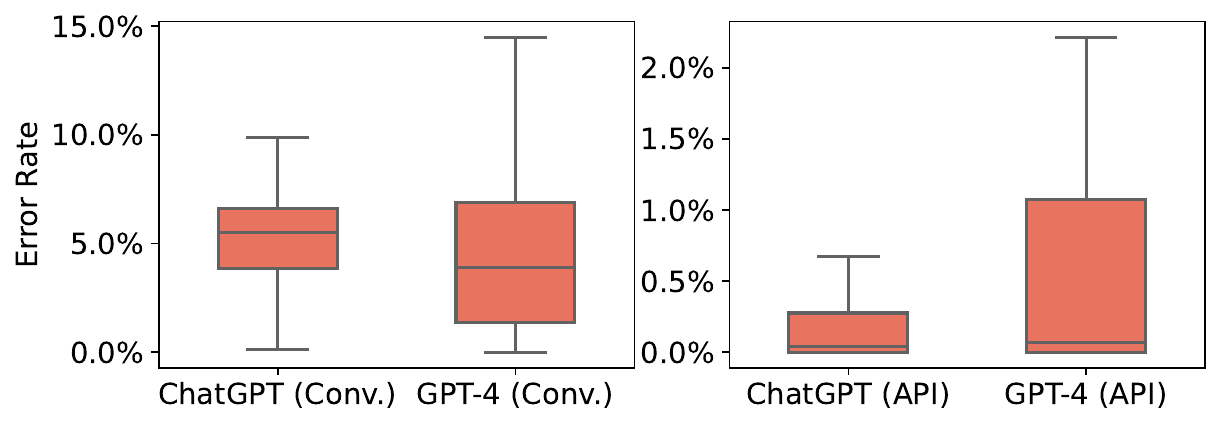}
        \vspace{-10pt}
        \captionsetup{font=footnotesize}
        \caption{Comparison of Statistical Failure Rates in \name.} 
        \label{fig:error}
    \end{minipage}
\end{figure}

We observe a high failure rate in GPT services. Figure~\ref{fig:error} shows that ChatGPT has a higher average service failure rate but better stability compared to GPT-4 in both conversational and API interfaces. Overall, GPT-4 has better service reliability over ChatGPT, demonstrating greater consistency and stability with fewer outliers. Additionally, the average failure rate of the conversation service is consistently high, exceeding 5\% for ChatGPT, which is significantly higher than that of regular cloud services~\cite{ahsan2023failure}. We will discuss the reliability problem in detail in Section~\ref{sec:result}.

% \begin{mdframed}[backgroundcolor=customblue]
% The API service demonstrates enhanced reliability compared to the conversational service. Considering the higher pricing of APIs, this disparity in performance may be attributed to the differential allocation of computational resources between the two services. The main reason behind the high failure rates of LLMs is worth investigating.
% \end{mdframed}

\section{BurstGPT-Perf: A Benchmark Suite}\label{sec:result}
This section presents our open-sourced benchmark suite: BurstGPT-Perf to evaluate LLM serving systems under \name~with streaming, stochastic, and bursty workloads. The suite is lightweight and modular, requiring minimal code integration for effective deployment. Additionally, it serves as a user-friendly example of how to implement \name~for evaluation and will be integrated and open-sourced within \name. With our use case, we advocate for using \name~on more serving systems to explore system limitations and optimization opportunities further.

\begin{figure}[htbp] 
\centering 
\includegraphics[width=0.45\textwidth]{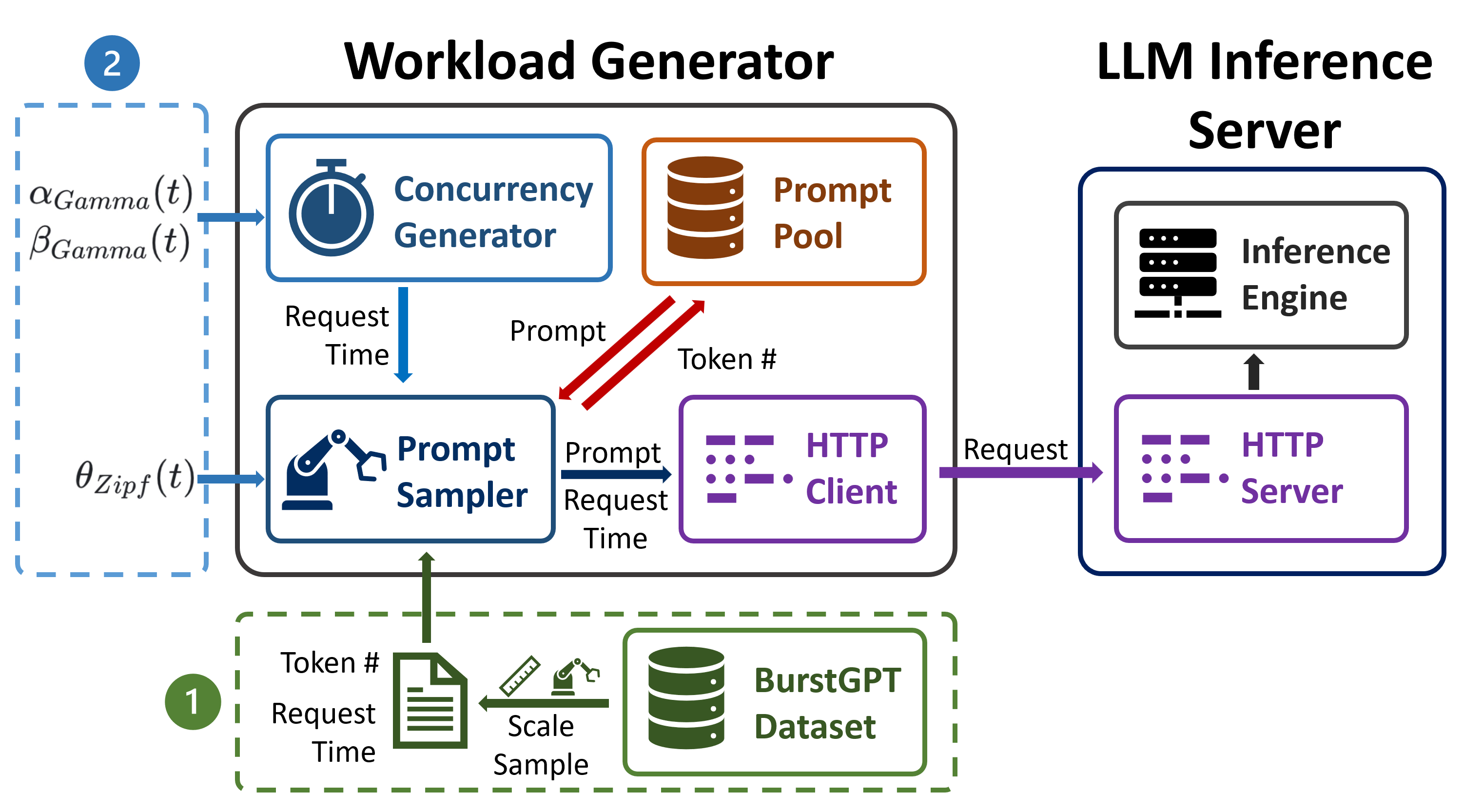} % new 
% \vspace{-20pt}
\caption{Overview of BurstGPT-Perf. It generates simulations of \name~in a burst manner with two scaling methods: 1. \textit{RPS Scaling} scales the original \name~data; 2. \textit{Modeled Scaling} uses Gamma distribution parameters to generate request times and Zipf distribution parameter to generate prompt token lengths.} 
\label{fig:workload_generator}
\end{figure}

% \begin{figure}[htbp] 
% \centering 
% \includegraphics[width=0.48\textwidth]{figures/workload_generator.png} % new 
% % \vspace{-20pt}
% \caption{Workload generator overview. It generates simulations of \name~in a burst manner. } 
% \label{fig:workload_generator_old}
% \end{figure}

\subsection{Implementation}
\paragraph{Workload Generator}
Figure~\ref{fig:workload_generator} presents an overview of the workload generator. The system comprises four main components: the \texttt{Prompt Sampler}, the \texttt{Prompt Pool}, the {Concurrency Generator}, and the \texttt{HTTP client}. The \texttt{HTTP} client is included using the efficient and asynchronous \texttt{aiohttp} framework. 
An index with token numbers as keys and prompt indices as values will be established in the \texttt{Prompt Pool} to mitigate the overhead of prompt sampler queries and a prompt is randomly selected for return. Once the prompt sampler retrieves a prompt from the \texttt{Prompt Pool} to enable workload concurrency, the concurrency generator times requests.
% This process concurrently cedes CPU control back to the prompt sampler, enabling it to proceed with the generation of the subsequent request. 
% In our implementation, the distribution parameters used by the prompt sampler and concurrency generator are time or request ID functions. 

\paragraph{Evaluation Workflow}

Requests of specified lengths are sampled from the \texttt{Prompt Pool} at stochastic intervals that are generated by \texttt{Concurrency Generator} in the \texttt{Workload Generator}. This concurrency pattern is treated as a time series, which inputs it into the \texttt{Inference Engine} via \texttt{HTTP} for performance evaluation purposes. The benchmark suite systematically adjusts RPS if we use the vanilla trace, or the parameters $\lambda$, $\alpha$, and $\beta$ according to a predetermined sequence, if we use modeled traces, thereby generating average and instantaneous performance metrics for users.

% \paragraph{Metrics Injection}
% We perform metrics injection at the LLM inference server for accurate and detailed metrics logging. Our metrics injection is conducted from two perspectives: request and engine step. For requests, we recorded the token number of each prompt and its response, as well as the arrival and completion time. Recognizing that the server might not always process requests immediately, we also recorded when the engine scheduler scheduled a request first. Considering scenarios where requests may be discarded due to server overload, it's essential to implement an auto-timeout feature within the LLM inference server. This functionality will automatically return a timeout response after a duration, effectively addressing the burstiness of real-world requests. From the engine step perspective, our records are more oriented toward hardware resource utilization and throughput data. We track each engine step's time, the number of tokens processed, and the GPU memory usage. 

\begin{figure*}[htbp]
\centering
\includegraphics[width=0.83\textwidth]{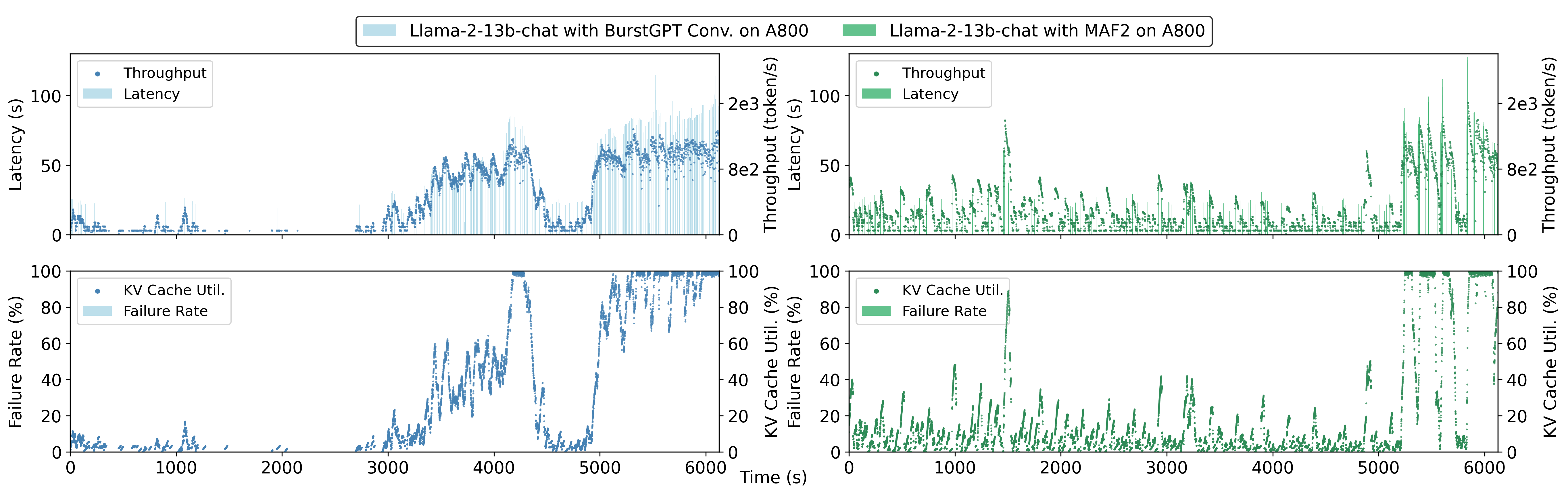}
\vspace{-10pt}
\caption{The impact of concurrency of \name~Conversation and \name~API traces on the latency, throughput, KV cache usage, and failure rates of vLLM serving on a single A800 GPU. Using the first 2000 queries for both traces, we scale BurstGPT by a factor of 10 to reach an RPS of 0.326. To reach the same RPS, MAF2 is scaled with a parameter of 1.234, ensuring both are complete in 6126 seconds.} 
\label{fig:stream}
\end{figure*}

\subsection{Scaling BurstGPT to Any Scale}

In Section~\ref{sec:trace}, we identify specific burstiness patterns from workload traces. To assess LLM serving systems on \name~at any scale, we scale \name~to match the system size, enabling scalable evaluations. We use \textit{RPS Scaling} or \textit{Modeled Scaling} in the evaluations. (1) For RPS scaling, we use a scaling parameter $c$ to multiply the timestamp, making the average RPS in the scaled trace equal to the given RPS. (2) Modeled scaling offers more flexibility in adjusting the scaling parameters.
Our analysis in Section~\ref{sec:trace} reveals that within \name, concurrency follows a varied Gamma distribution, characterized by the shape parameter $\alpha$ and the scale parameter $\beta$. The parameter $\alpha$ predominantly influences the coefficient of variation ($CV$) of the distribution, mathematically expressed as $CV = 1/\sqrt{\alpha}$.

To make \name~applicable for evaluating LLM serving systems of different sizes, we adjust these parameters in \name's concurrency distribution model at intervals of 20 minutes. For assessments of \name~on a serving system of a particular scale, users can modify the $\beta$ parameter to determine a warm-up arrival rate compatible with their system's capacity, such as 30\% to 40\% GPU utilization of KV cache in our setting. Following this adjustment, users can evaluate their system within any chosen time range in \name.

% \paragraph{Self-defined Variation}
% The users can also define the concurrency variation to simulate different burstiness shift patterns.
% In the initial setting, evaluations employ a quadratic function to model the variable burstiness of $\alpha$ and a linear function for the variable burstiness of $\beta$. Using a quadratic function for shape parameter $\alpha$ allows for capturing rapid, non-linear variations in burstiness observed in Section~\ref{sec:trace}. Conversely, using a linear function for $\beta$, the scale parameter, aligns with its role in controlling the distribution's spread, suggesting more gradual changes. This approach addresses the distinctive influences of each parameter on the system's behavior under varying burstiness.

\subsection{Matching BurstGPT with Dialogues}
For request lengths, in Section~\ref{sec:trace} and related simulations~\cite{li2023alpaserve}, it is observed that the request length follows a Zipf distribution. The sampler samples request lengths using a variable parameter $\lambda$, influencing the frequency of shorter requests in input streams.

For response lengths in \name, they are not predefined, allowing the model to determine them dynamically. This strategy avoids restrictions that could degrade system runtime, especially with specific LLMs. Over many requests, the distribution of output lengths will naturally align with the model's behavior, as detailed in Section~\ref{sec:trace}.

After determining request lengths, the suite initializes a CPU buffer for loading a truncated prompt dataset pool. This dataset can be sampled from existing GPT conversation dialogues, such as Alpaca~\cite{li2023alpaca}. The dataset truncation and request length sampling adhere to the predefined Zipf distribution. The buffer size is user-configurable to balance with the CPU's memory capacity. Requests of specific lengths are then selected from the prompt dataset and sent to the server at intervals, where they queue for processing.

\subsection{Metrics and Setups}
In serving systems, Quality of Service (\textit{QoS}) is often measured by \textit{latency} and \textit{throughput}~\cite{miao2023efficient}. Latency refers to the completion time of each token, while throughput is the number of tokens processed within a given period.
In addition to efficiency, the stability and reliability of LLM serving are particularly important to users, yet these aspects are often overlooked. In this evaluation, we also focus on latency jitters to assess stability and failure rates to assess the reliability of LLM serving.

We categorize metrics into two types: average values (denoted with a superscript $^{avg}$) and instantaneous values (marked with a superscript $^{ins}$). This categorization provides a comprehensive understanding of both long-term trends and immediate system behavior. Based on the trace study, the following key metrics are employed for a thorough evaluation of a serving system's performance:
\begin{itemize}
\item $R^{avg}$ / $R^{ins}$: Request Failure Rate;
\item $L^{avg}$ / $L^{ins}$: Token Latency;
\item $\sigma_L^{avg}$ / $\sigma_L^{ins}$: Token Latency Jitters (Standard Deviation);
\item $P^{avg}$ / $P^{ins}$: System Throughput.
\end{itemize}

\section{Demo Evaluations}\label{sec:case}
Our evaluation is conducted on Llama-2-13b-chat on an A800 and A6000 GPU server. In experiments that use modeled scaling, we set the initial parameters $\alpha_{Gamma}$ and $\beta_{Gamma}$ of the Gamma distribution in \name~to 0.5 and 2, respectively, for 30\% GPU KV Cache utilization. The request length parameter $\theta_{zipf}$ is set to 1.1. We use requests in ShareGPT as the request pool.

\subsection{Evaluation: BurstGPT v.s. MAF2}

\begin{figure}[htbp]
\begin{center}
% \includesvg[width=1\linewidth]{figs/time_cost.svg}
\includegraphics[width=0.7\linewidth]{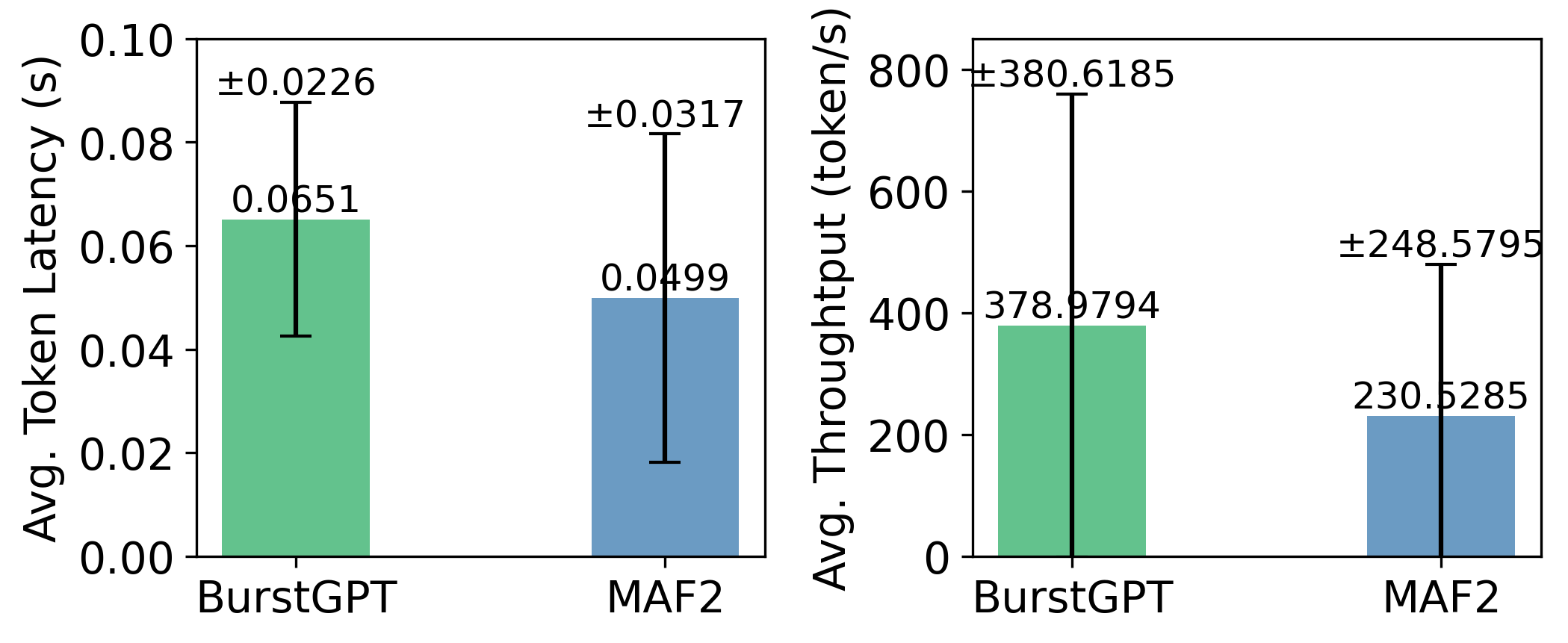}
% \vspace{-15pt}
\caption{Average token latency and throughtput of vLLM serving Llama-2-13b-chat on an A800 GPU server with \name~and MAF2 workloads.}
% \vspace{-20pt}
\label{fig:bgpt_mafg_2pic}
\end{center}
\end{figure}

In this evaluation, we assess the efficiency and stability of vLLM~\cite{vllm} using \name~and MAF2. We set the same RPS to 0.326 to scale both MAF2 and \name. Both experiments were completed in 6126 seconds. Figure\ref{fig:stream} shows the serving system's instantaneous metrics under the two workloads. The results indicate that the request burstiness of concurrency in \name~is higher, whereas the request distribution of the MAF2 workload is more uniform.  As we can see from Figure \ref{fig:bgpt_mafg_2pic}, the average latency and throughput are higher for \name~compared to MAF2, indicating a larger system resource occupation by \name.
In conclusion, although both methods share the same RPS, the more bursty concurrency in \name~introduces more challenges to the serving system, particularly in terms of KV cache management, compared to MAF2.

% We deploy Llama-2-13b-chat model via vLLM on a single A800 GPU server, and separately evaluate this serving system using \name~and MAF2 as workloads. We sample the first 2000 queries for both workloads. (Why set the scale to 10?) After some small scale experiments using different scale parameter, we set the scale parameter to 10 in \name, which will make the workload 0.326 RPS. To reach the same RPS, the system adjusts the scale parameter in MAF2 workload to 1.234.

\subsection{Evaluation: Micro Workload Variations}

\begin{figure*}[t] 
\centering
\includegraphics[width=0.83\textwidth]{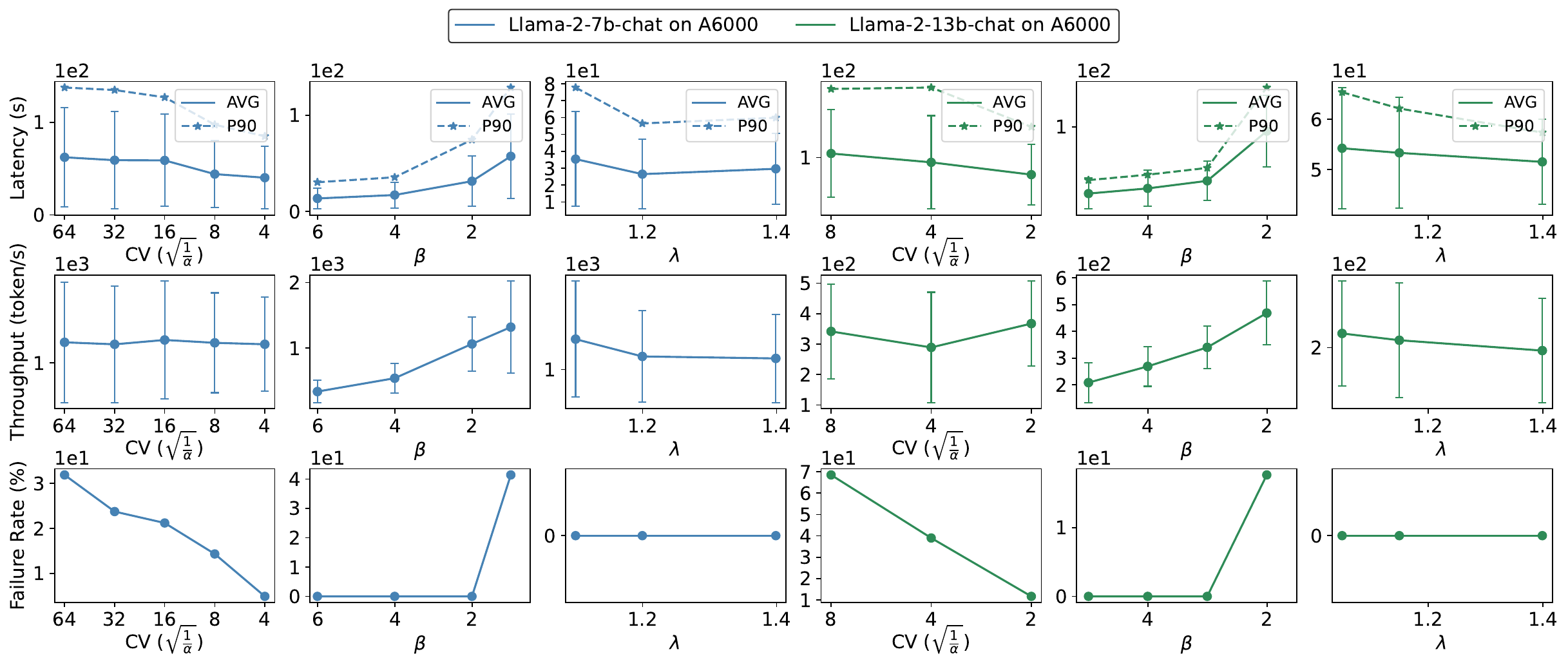} 
\vspace{-10pt}
\caption{Impact of parameters $\alpha$, $\beta$, and $\lambda$ on average and p90 latency, average throughput, and average failure rates in vLLM. The evaluation is conducted with Llama-2-7b-chat and Llama-2-13b-chat models using A6000 and A800 GPU Servers. Experiment Settings: Larger $\alpha$ or smaller $CV$ and $\beta$ increase system burstiness; higher $\lambda$ correlates with shorter request likelihood. Note: Due to space limits, only results from the A6000 server are shown.
} 
\label{fig:final}
\end{figure*}

In this experiment, we model the request lengths to a Zipf distribution, as in~\cite{li2023alpaserve}. The sampler samples the request lengths in Zipf distribution with a variable parameter $\lambda$. We investigate the impact of varying specific variables: $\alpha$, $\beta$, and $\lambda$. We utilized the Llama-2-7b-chat and Llama-2-13b-chat models in our testbed. The results demonstrate a correlation between the adjustments in these variables and the observed system behavior.

We observe that minor changes in the parameter $\alpha$ or the $CV$ can lead to sudden request failures, significantly undermining the system's reliability. However, these changes have less effect on system performance metrics such as average latency and throughput. This is because such changes in $\alpha$ are temporary and behave as quadratic functions. In contrast, alterations in $\alpha$ in a linear decline lead to ongoing performance degradation, particularly noticeable during request failures. These findings underscore the value of our trace analysis and provide essential insights for improving system efficiency in similar computational environments.

\subsection{Demo Use: Request Scheduler Selection}
\begin{figure*}[htbp]
\centering
\includegraphics[width=0.75\textwidth]{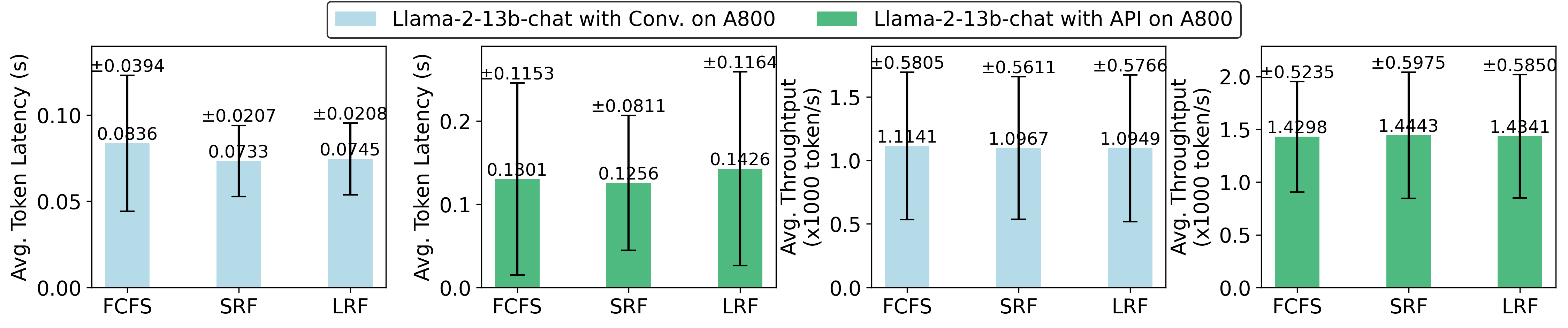}
\caption{Average token latency and throughput of FCFS, SRF, and LRF on vLLM serving Llama-2-13b-chat on an A800 GPU server with \name~Conversations and API workloads.} 
\label{fig:avgtokenlat}
\end{figure*}
We randomly sample a time range on \name~and build two subsets with conversation workloads and API workloads and then used Modeled Scaling to derive the corresponding parameter sequences for generating workloads. We use these two generated workloads separately to evaluate three scheduling strategies i.e. fisrt-come-first-serve (FCFS), short-request-first (SRF), and long-request-fisrt (LRF), on vLLM. 

% For FCFS, in Figure \ref{fig:stream2}, the failure rate is low on conversation trace, whereas it turns high due to the extreme burstiness in API service. 
In Figure~\ref{fig:avgtokenlat}, the efficiency of the various scheduling methods may shift across different types of real-world workloads.
Comparing FCFS with SRF and LRF on average values in terms of efficiency and stability, we can find that the optimization on conversation service when LRF is superior to FCFS in latency and and latency jitters does not stand in API service. This interesting finding inspires us that even using real-world workloads,
an optimization on one workload does not necessarily generalize to other workloads.

\subsection{Demo Use: Workload Provisioning}
% BurstGPT plays an important role in facilitating the study of user and model behaviors for real-world LLM services, enabling the development of an efficient and robust scheduling system. In this section, we present a simple toy example to demonstrate BurstGPT's potential for research aimed at improving real-world service deployment.

% \paragraph{Provisioning Algorithms}
\label{sec:load_prediction}

\begin{figure}[htbp]
\centering
% \hspace{-12px}
\includegraphics[width=.9\linewidth]{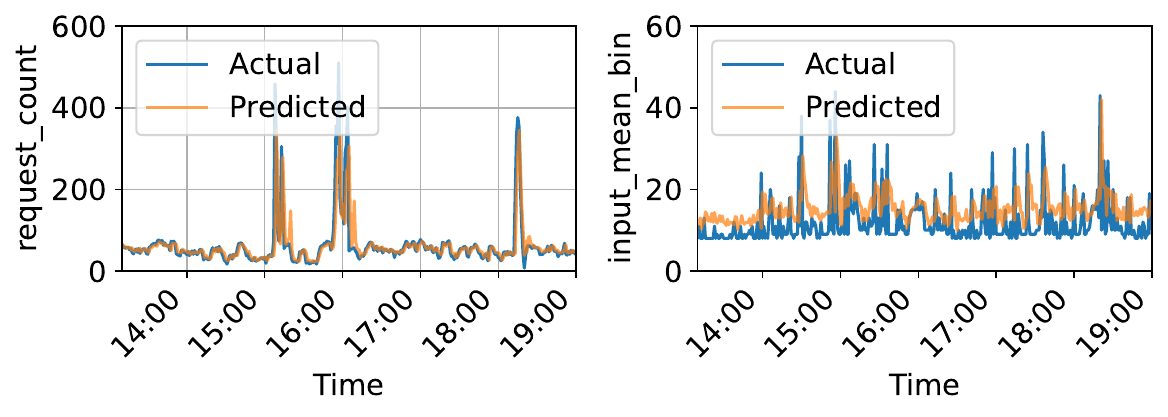}
\caption{Request count and mean request token number predictions. The figure only shows a random interval of 6 hours. The time bin interval is 1 minute. Left: Request count predictions, the NMAE is 0.73 and the NMSE is 0.48. Right: Mean request token number predictions, the NMAE is 0.72 and the NMSE is 0.67.} 
\label{fig:pred_all}
\end{figure}

To adjust the resources allocated to services in a timely manner, cloud systems can predict system load to proactively scale resources up or down.
The workload provisioning task is a time series forecasting problem. Specifically, we select the request count per time bin and the average number of tokens per request as the prediction targets. To construct features for the prediction model, we use historical load data from both short-term and long-term perspectives. For each time point, we incorporate load values from the past 3 time points (lag window) to capture short-term trends, along with statistical features (e.g., mean, variance) computed over a rolling window of up to 60 minutes to reflect long-term patterns. As shown in \ref{sec:Long-termPatterns}, the day of the week and the hour of the day also impact the workload, so these two features are included. The prediction target is the actual load value for that time point. We employ XGBoost~\cite{chen2016xgboost} as the prediction model due to its simplicity and effectiveness. We compute the normalized mean absolute error (NMAE) and normalized mean square error (NMSE) between the prediction and the ground truth. The results are shown in Figure~\ref{fig:pred_all}.

As shown, even without careful tuning, the models can predict future system load easily and accurately. The predicted load trends generally align well with the actual load, indicating significant opportunities for fine-grained load prediction modeling in real-world inference services, which lead to more precise service delivery.

\begin{figure}[htbp]
\centering
\includegraphics[width=.9\linewidth]{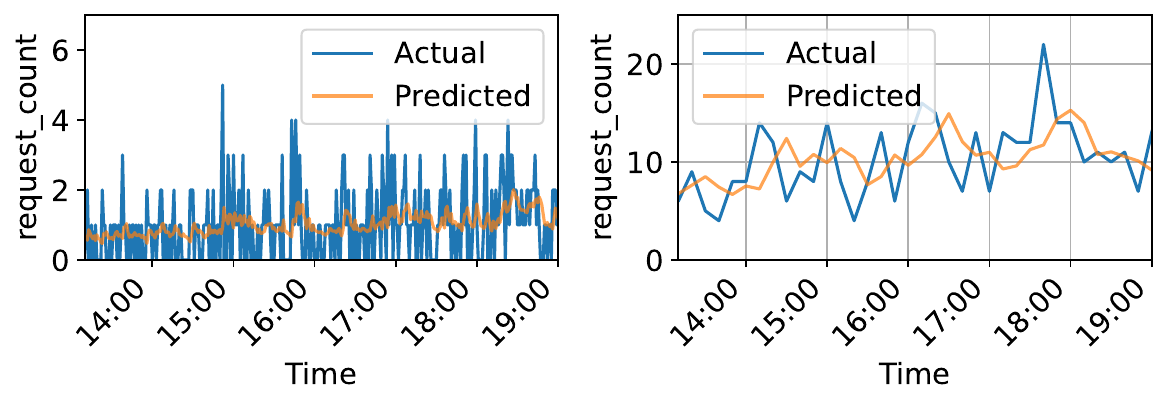}
\caption{Request count predictions. The figure only shows a random interval of 6 hours. Left: Request count predictions of Conv. with time bin interval 1 minute, the NMAE is 0.66 and the NMSE is 0.50. Right: Request count predictions of Conv. with time bin interval 10 minutes, the NMAE is 0.32 and the NMSE is 0.21.} % for conv only
\label{fig:pred_conv}
\end{figure}

We also investigate the performance of the prediction algorithm under different time intervals. As shown in Figure~\ref{fig:pred_conv}, predicting long-term patterns (10 minutes) is generally easier than predicting short-term patterns (1 minute), with the former achieving a 58\% lower NMSE compared to the latter. This suggests that when using load prediction for serving, it is essential to balance the granularity of scheduling with the accuracy of predictions.

% BurstGPT generally provides both industry and academia with real-world data to study and validate algorithms related to load prediction.

% \paragraph{Scheduling under Provisioned Workload}

% To further demonstrate the advantages of the scheduling algorithm using data-driven load prediction, we compare it against a naive scheduling algorithm that does not. Specifically, the naive scheduling algorithm monitors the KV cache usage of serving instances in real-time and initiates/kills \textcolor{blue}{x} additional workers when the KV cache usage exceeds/falls below predefined thresholds \textcolor{blue}{x and y} respectively.
% For the prediction-based scheduling, we adopt the XGBoost model described in Sec. \ref{sec:load_prediction}. It will compute the projected KV cache usage given predicted input and output loads, and dynamically scale the number of instances accordingly.


\section{BurstGPT in Industry}
We provide an example of how we use \name~for industrial use in Huawei during our development of $PD$ disaggregation systems. As a common practice in the industry, given a fixed serving system scale, developers use goodput~\cite{distserve} to evaluate the system's capacity, i.e., the maximum request rate that can be served while meeting the SLO goal (e.g., 90\%) for each GPU. Recently, disaggregating the prefilling and decoding processes during LLM inference has become the standard approach in LLM service deployment~\cite{memserve,kimi,distserve}. However, scheduling the disaggregation of prefilling and decoding can be challenging. The prefilling instances can be seen as the "Provider," which generates the first KV cache and token, while the decoding instances act as the "Consumer," accepting the KV cache and token to continue the auto-regressive generation. Thus, for a given system scale (i.e., the number of prefilling ($P$) and decoding ($D$) instances), the concurrency of requests, request lengths, and response lengths will determine the goodput of the system. As workloads vary rapidly in real-world serving, a fixed $PD$ ratio is not always ideal. Instead, a dynamic $PD$ ratio should be used.

\begin{figure}[htbp]
\centering
\includegraphics[width=0.7\linewidth]{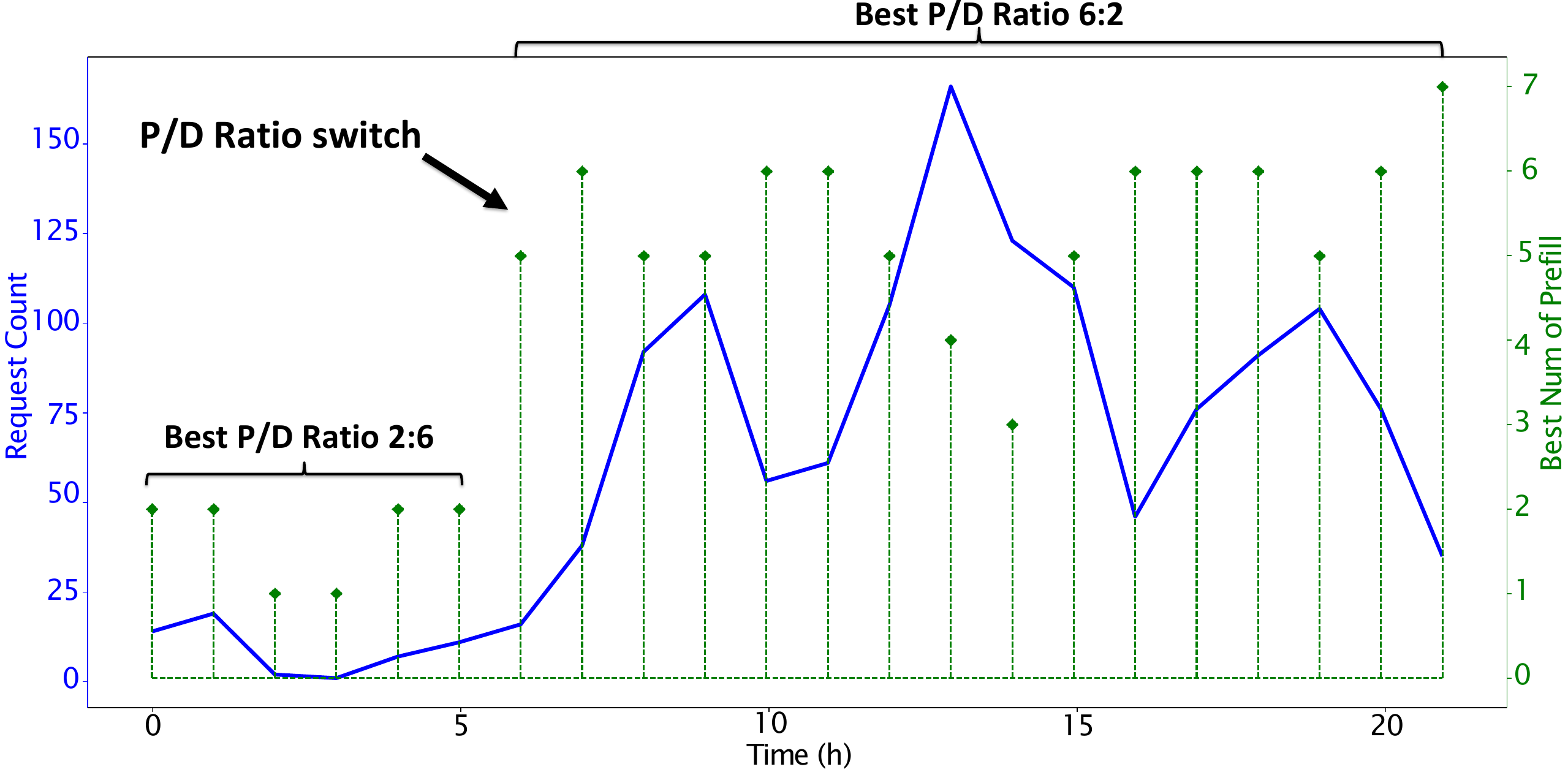}
\caption{At 6:00, the PD ratio switches from 2:6 to 6:2, improving the goodput of the serving system.} 
\label{fig:goodput2}
\end{figure}

\begin{figure}[htbp]
\centering
\includegraphics[width=0.7\linewidth]{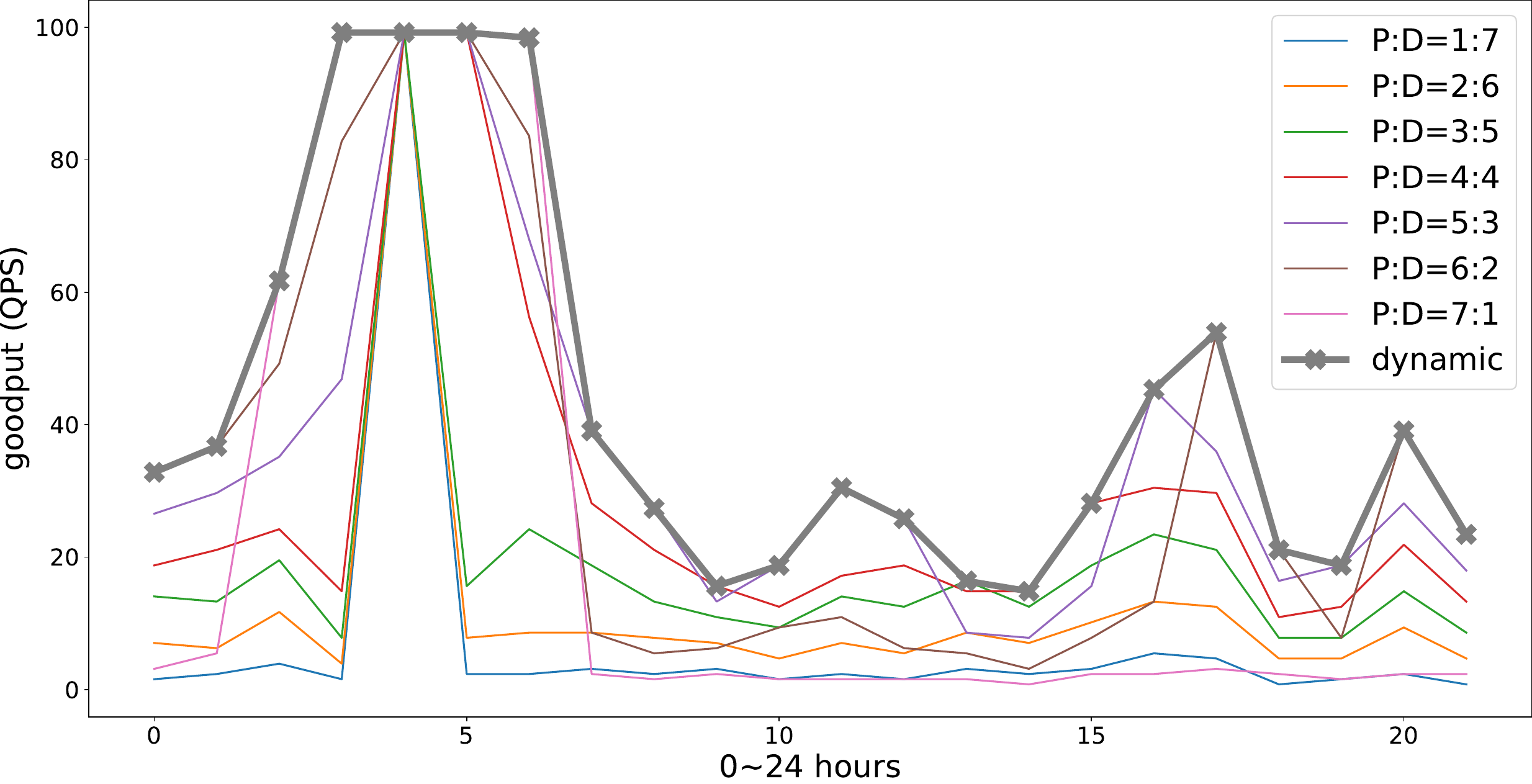}
\caption{Given 8 instances, comparison of different PD ratio and dynamic PD ratio based on workload provisioning and beam search during LLM serving.} 
\label{fig:goodput1}
\end{figure}

Figure~\ref{fig:goodput1} and Figure~\ref{fig:goodput2} show examples of goodput at different PD ratios. Since we lack preliminary workload traces before the real-world deployment of our serving system, we use \name~in our simulation platform to determine the optimal PD ratio under various user, model, and system workloads. The result is used to develop prototypes of the $PD$ disaggregation serving system at Huawei. In the simulation, we select the $P$ and $D$ instance ratio, then run simulations, gradually increasing QPS while monitoring the delay SLO achievement rate until the SLO is no longer met. The QPS at this point is considered the system's capacity. We then fix the total number of $P$ and $D$ instances. 
First, considering a simple situation, we adjust the $PD$ ratio one time to fit the workload divergence in Figure~\ref{fig:goodput2}, we observe that at 6:00, the $PD$ ratio switches from 2:6 to 6:2, improving the goodput of the serving system. This simulation serves as a solid prototype for real-world $PD$ disaggregation scheduling policies. 
Then, we develop a dynamic $PD$ ratio strategy. We use beam search to identify the best $P$$D$ ratio given the provisioned workload in BurstGPT. Instead of directly scaling the number of instances, we simulate scaling by adjusting the load size per instance, assuming a predefined instance launch time in seconds. As shown in Figure~\ref{fig:goodput1}, with the provisioned workload, the system searches for the optimal $PD$ ratio to improve the goodput. 

These findings highlight the significant improvements in LLM serving achieved through data-driven modeling of user and model behavior. By carefully designing the scheduling algorithm, we can greatly enhance both system efficiency and QoS.

\section{Broader Impacts and Conclusions }\label{sec:impacts}
This research emphasizes incorporating real-world workload data to enhance LLM serving systems. It identifies a critical gap: the need for such data in optimizing these systems. In this work, we introduce the real-world LLM workload \name~from Azure OpenAI GPT service. The traces and evaluations reveal possible performance degradation of serving systems under real-world workloads, highlighting challenges and opportunities in workload-aware LLM serving optimizations.

We encourage using \name~in optimizing and evaluating serving systems to improve the efficiency, stability, and reliability of LLM services under realistic workloads. We also advocate for data-driven methodologies in developing LLM systems in the future.

\section{Acknowledgement}
This work that involves all hardware use was done and open-sourced while Yuxin Wang was a research intern at the Hong Kong University of Science and Technology (Guangzhou). It was partially supported by the National Natural Science Foundation of China under Grant No. 62272122, a Hong Kong RIF grant under Grant No. R6021-20, and Hong Kong CRF grants under Grant No. C2004-21G and C7004-22G. Also supported by the the High-performance Computing Platform and the Red Bird MPhil Program of the Hong Kong University of Science and Technology (Guangzhou). For all closed-source models, evaluations are performed through their respective APIs.

\bibliographystyle{ACM-Reference-Format}
\bibliography{sample-base}

\end{document}